\documentclass[english]{revtex4-1}
\usepackage[T1]{fontenc}
\usepackage[latin9]{luainputenc}
\usepackage{graphicx}

\makeatletter
 \@ifundefined{textcolor}{}
 {%
   \definecolor{BLACK}{gray}{0}
   \definecolor{WHITE}{gray}{1}
   \definecolor{RED}{rgb}{1,0,0}
   \definecolor{GREEN}{rgb}{0,1,0}
   \definecolor{BLUE}{rgb}{0,0,1}
   \definecolor{CYAN}{cmyk}{1,0,0,0}
   \definecolor{MAGENTA}{cmyk}{0,1,0,0}
   \definecolor{YELLOW}{cmyk}{0,0,1,0}
 }

\makeatother

\usepackage{babel}
\begin{document}

\title{A simple Cherenkov detector for educational purposes}

\author{S. Ozturk}

\address{Gaziosmanpasa University, Physics Dept., Tokat, Turkey}

\author{G. Unel}

\address{University of California at Irvine, Physics and Astronomy Dept.,
Irvine, CA, USA}

%\author{B. Bilki}

%\address{Beykent Univ., Physics Dept, Ankara, Turkey}
\begin{abstract}
Cherenkov detectors are in use today in small experiments as well
as modern ones as those at the LHC. This short note is about the construction
of a small Cherenkov detector with limited resources, which could
be used to observe the cosmic rays. The Cherenkov light obtained in
this particular detector is read with a photo multiplier tube and
its signal is observed on an oscilloscope. The detector construction
can be achieved by relatively simple means and can be used as an educational
tool.
\end{abstract}
\maketitle

\section{Introduction}

\begin{figure}
\begin{centering}
\includegraphics[width=0.4\textwidth]{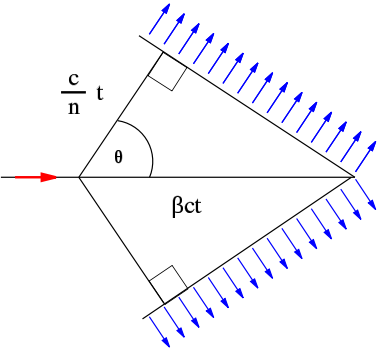} $\quad$\includegraphics[width=0.4\textwidth]{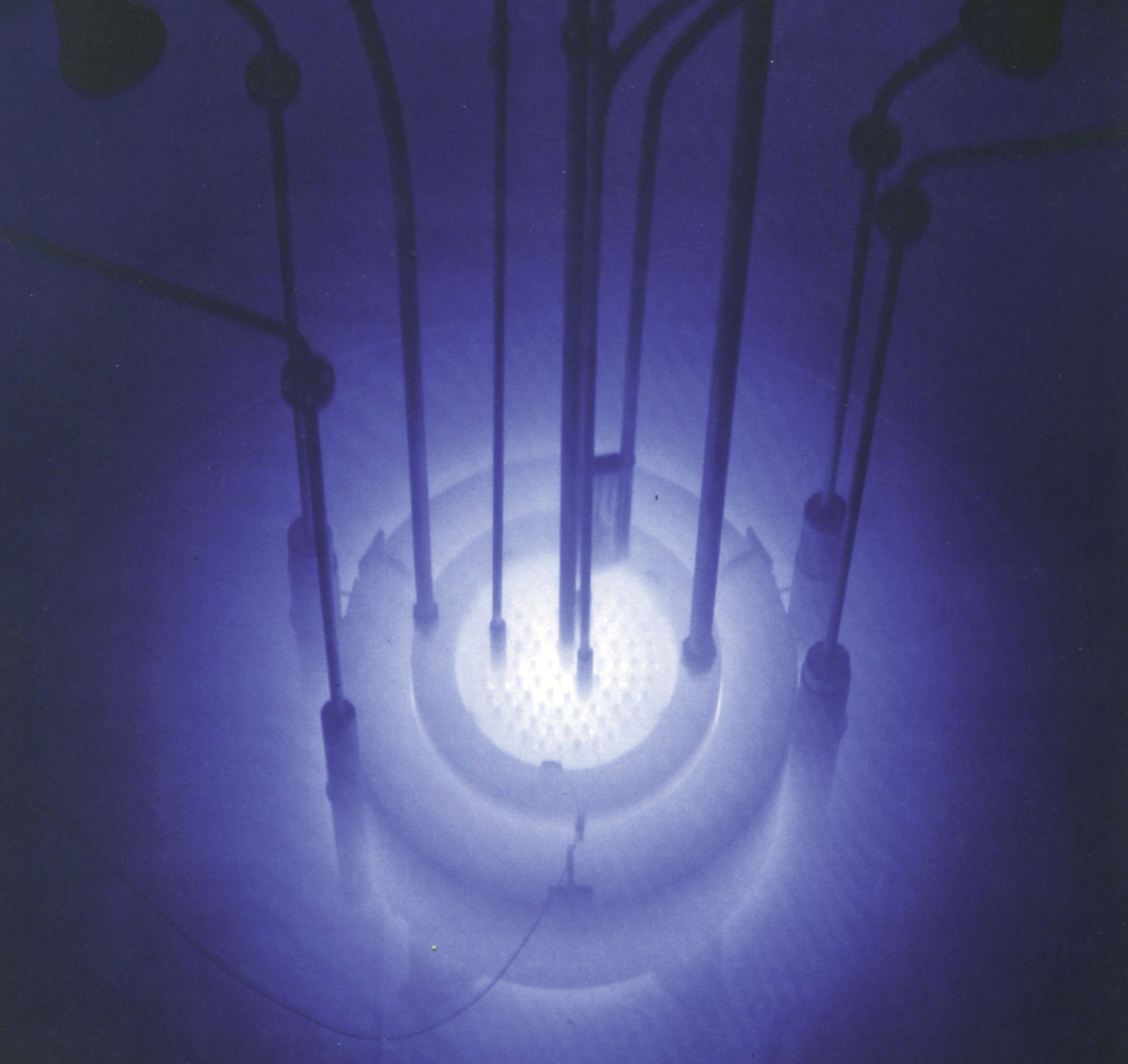}
\par\end{centering}
\caption{Cherenkov light production mechanism and geometry on the left and
Cherenkov light in the Reed Research Reactor on the right \label{fig:cherenkov-typical}}
\end{figure}

Cherenkov radiation is electromagnetic radiation when a charged particle
passes through in a dielectric medium at a speed greater than the
speed of light in that medium. When a charged particle travels, it
disturbs the electric field in the medium and medium becomes electrically
polarized. If a charged particles has enough speed, a coherent shockwave
is emitted. The similar analogy can be though as a sonic boom of a
supersonic aircraft. Cherenkov emitting angle can be easily expressed
as $cos(\theta_{c})=\frac{1}{n\beta}$ , where $n$ is the refractive
index of the material and $\beta$ is $v/c$. If $\beta$ is greater
than $1/n$, Cherenkov radiation will be emitted as shown in Fig.\ref{fig:cherenkov-typical}
left side. Using relativistic kinematics, the threshold energy of
Cherenkov radiation can be given by: $E_{threshold}(n,m)=mc^{2}\frac{n}{\sqrt{n^{2}-1}}$
\cite{thesis}. At this energy, the Cherenkov light is emitted along the
path of the particle. The same figure right side contains the blueish
Cherenkov light observed at the Reed Research Reactor originating
from fast electrons. In case of $\Delta n=n-1\ll1$ approximation,
the threshold energy can be rewritten as $E_{threshold}(n,m)=\frac{mc^{2}}{\sqrt{{2\Delta n}}}$
\cite{thesis}. Therefore, such a detector could be used to trigger on various
particles at various energies for small experiments or for educational
purposes, depending on the radiation medium.

\section{Detector Design and Construction}

The simplest Cherenkov detector would consist of a barrel with enough
volume to produce Cherenkov light and reflective inner surfaces to
channel that light into photon counting detectors. In high energy
physics such a detector would be a Photo Multiplier Tube (PMT) \cite{pmt0} requiring
about 1.5-2 kV for operation. The speed of light in the material to
be installed into the barrel has to be smaller the vacuum value (\emph{c})
to permit Cherenkov radiation. For example water ($v_{\gamma}=0.75c$)
would be the a cheap and easily accessible material allowing detection
of Cherenkov light for cosmic muons with energy larger than 160 MeV.
However using CO2 (Air) would allow detection of cosmic muons of energy
threshold of 3.52 (4.4) GeV.

\begin{figure}
\begin{centering}
\includegraphics[width=0.4\textwidth]{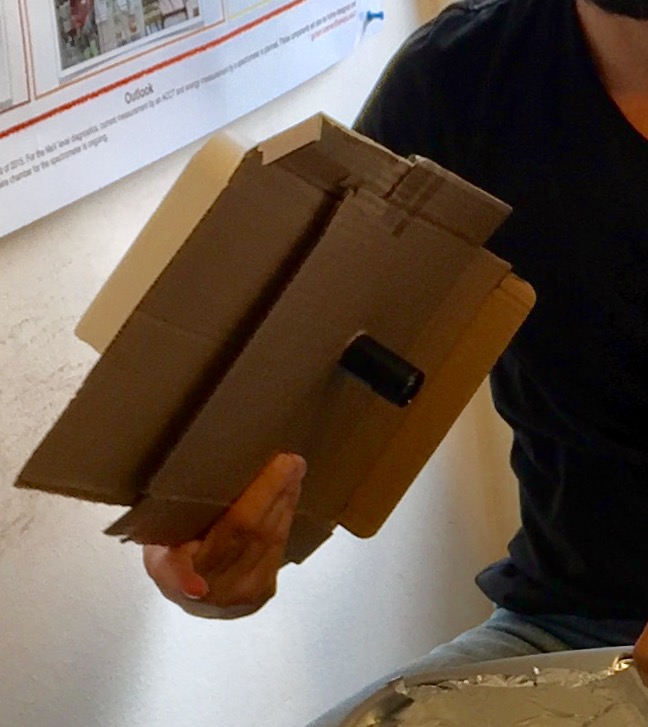} 
\par\end{centering}
\caption{A 20L dust bin with inner surfaces covered with aluminum foils was
used as the vessel for Cherenkov light producing material: water.
\label{fig:the-bin}}
\end{figure}

To build the simplest possible detector in the cheapest possible way,
possible barrel alternatives have been considered such as a beer keg
and a large water dispenser. The simplest solution was found to be
a 20L volume plastic trash bin since it permits about 50 cm water
depth, large enough to produce Cherenkov light. The inner side of
the bin was covered with kitchen grade aluminum foils to improve its
reflectivity. A view of the \emph{Al} covered bin can be seen in Fig.
\ref{fig:the-bin}. The available PMTs were R7525HA-2 from Hamamatsu
\cite{pmt} with 25mm photocathode diameter and requiring about 1500V
for proper operation. Since these PMTs can not operate
in water, a simple setup from foam and cardboard was prepared to keep
the single PMT above the water level. The signal output from the PMT
base can be readout directly with an oscilloscope without needing
a pre-amplifier circuit. The light tightness of the finished product
was provided by multiple layers of black felt blankets.

%\begin{figure}
%\begin{centering}
%\includegraphics[width=0.4\textwidth]{figs/IMG_20170815_153437} $\quad$\includegraphics[width=0.37\textwidth]{figs/IMG_20170815_155308}
%\par\end{centering}
%\caption{Cabled setup and the light tight detector }
%\end{figure}

\section{Detector Tests}

\begin{figure}
\begin{centering}
\includegraphics[width=0.6\textwidth]{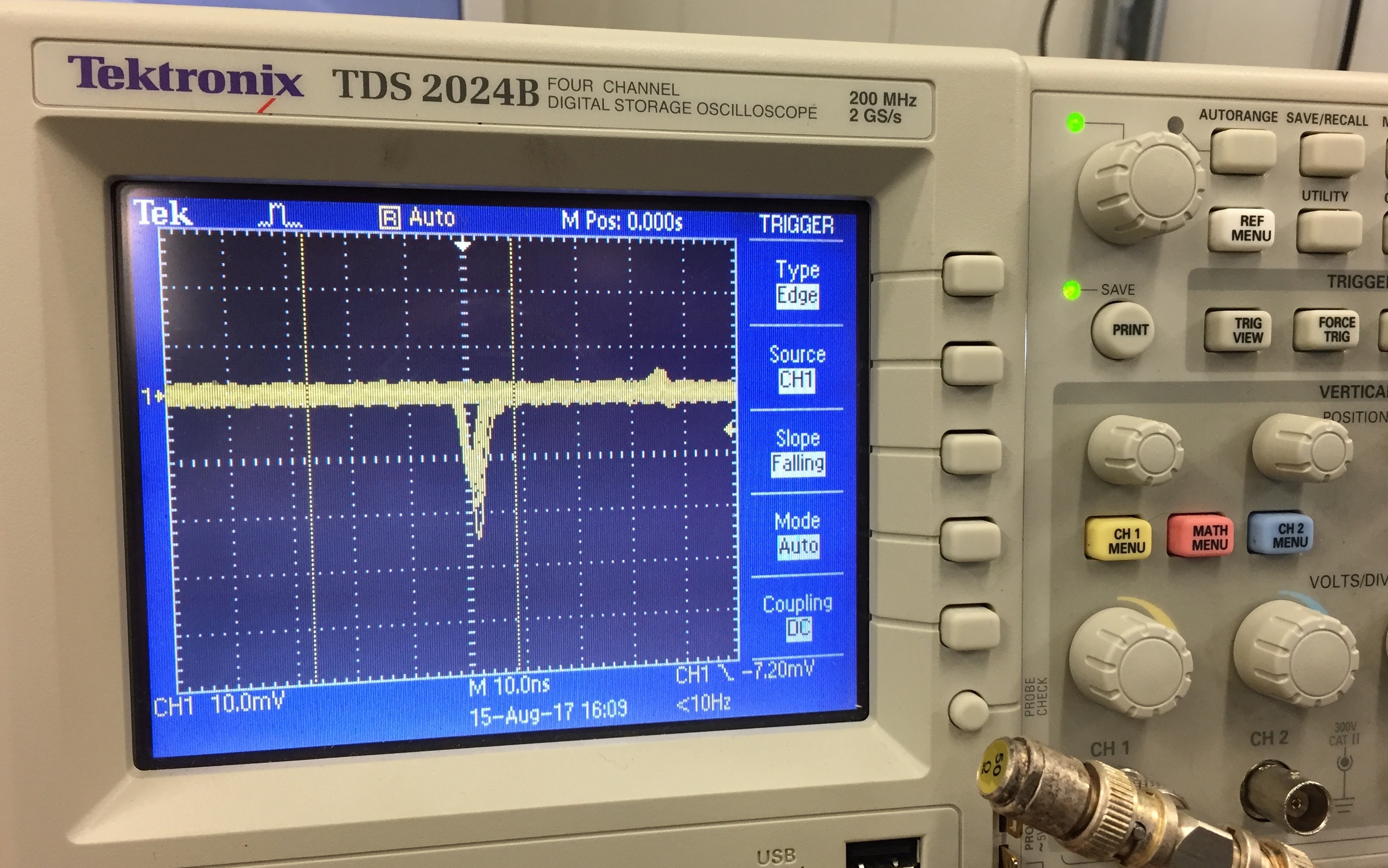}
\par\end{centering}
\caption{Background measurements with air in the bin\label{fig:Backgrd} }
\end{figure}

The detector was first tested with air inside the bin to have an understanding
of the background values. With a threshold as low as -7.2 mV, the
background rate is measured to be less than 10Hz. This value is consistent
with a background rate originating from the PMT's own dark current
and from cosmic rays passing through the PMT window. A screenshot
from the oscilloscope in accumulating display mode can be seen in
Fig. \ref{fig:Backgrd}. 

\begin{figure}
\begin{centering}
\includegraphics[width=0.37\textwidth]{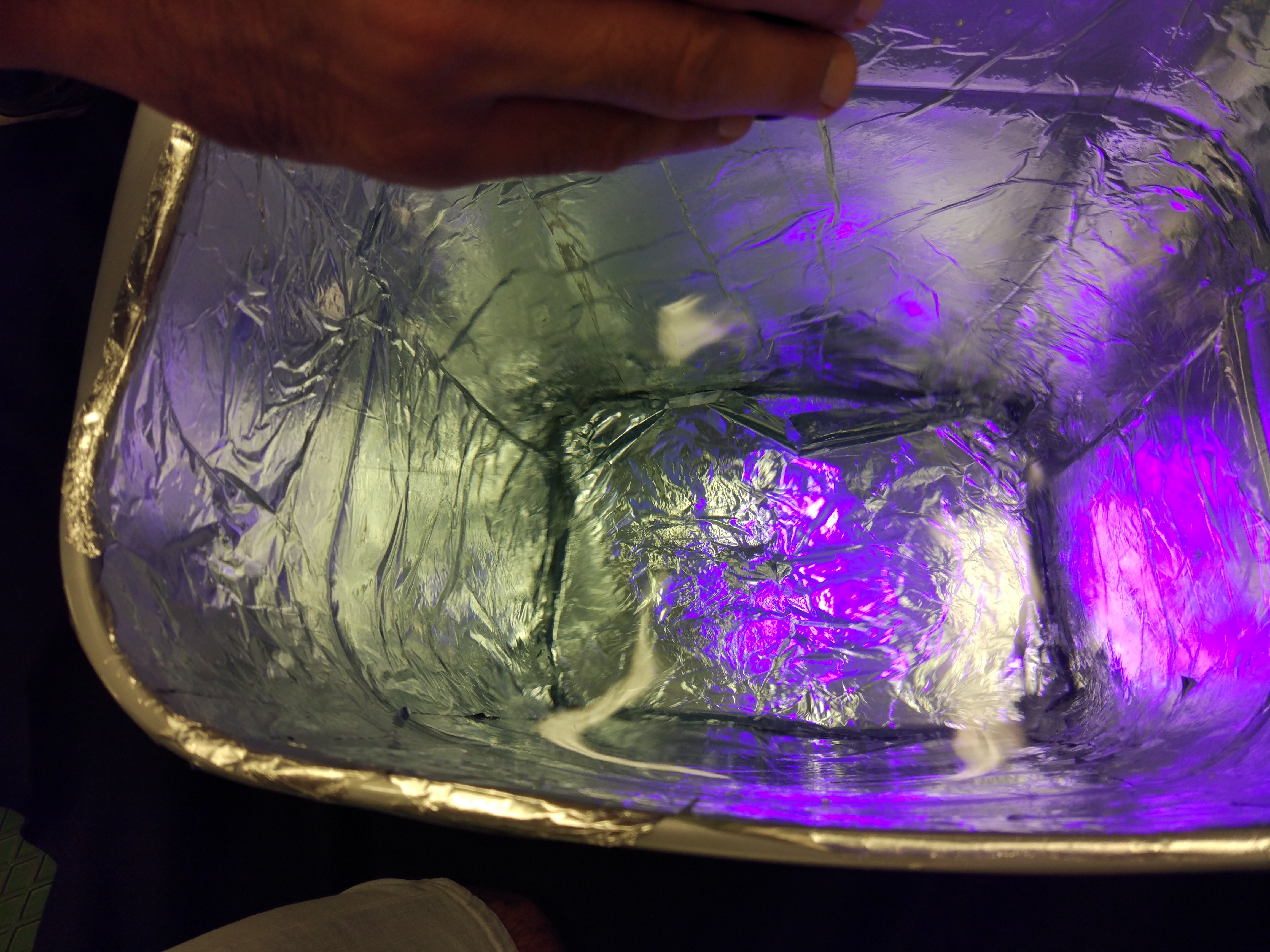}$\quad$\includegraphics[width=0.4\textwidth]{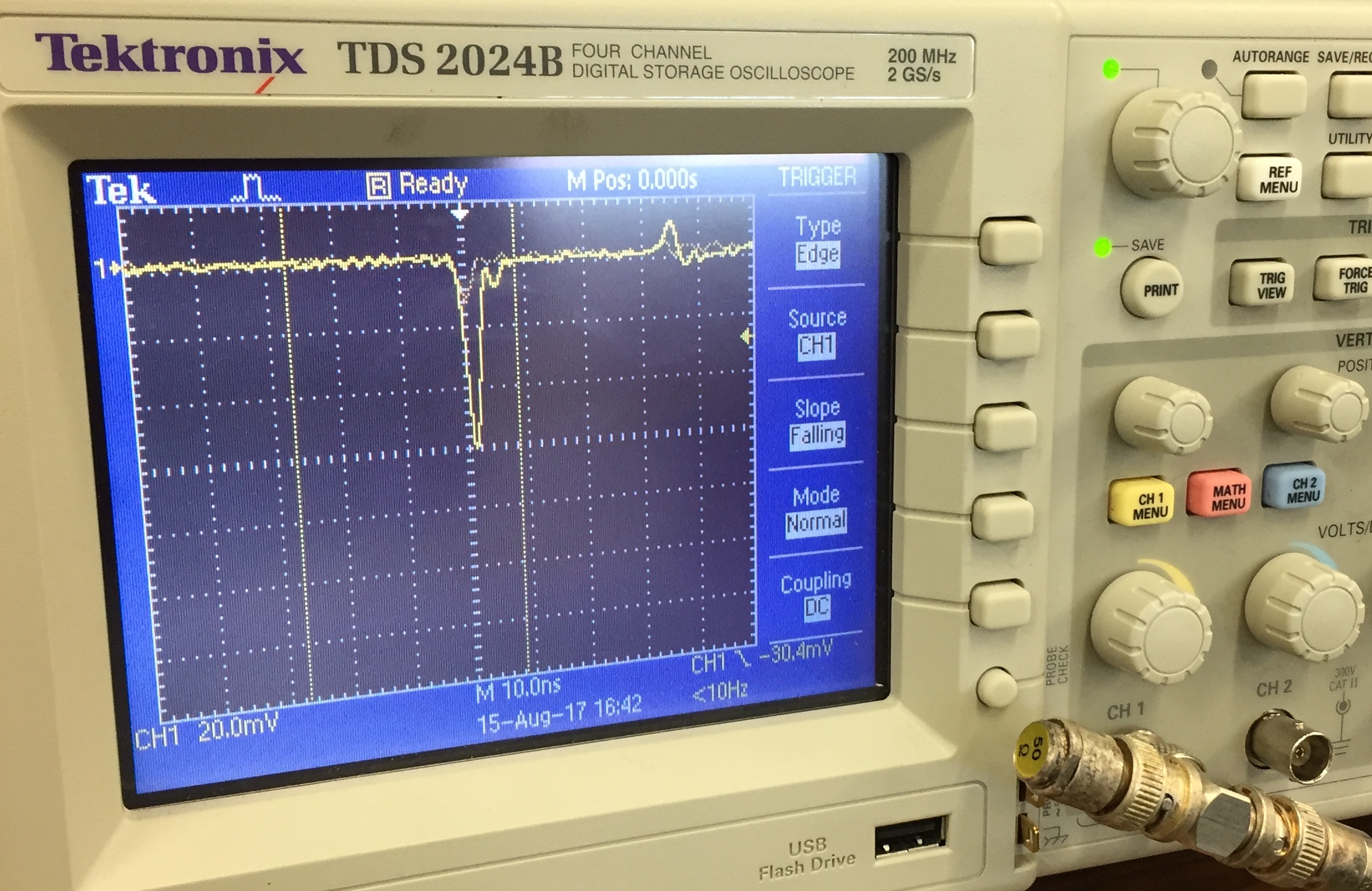}
\par\end{centering}
\caption{Background measurements with water in the bin\label{fig:Background-water} }
\end{figure}

After determination of the rate and pulse hight of the background
events, the bin was filled with tap water providing a water height
of about 50cm as seen in Fig.\ref{fig:Background-water}left side.
The signal threshold was set to -30mV eliminating most of the backgrounds.
The solid line pulse in the same figure right side acquired after
about 10 mins is consistent with a cosmic ray passing through the
photocathode, both rate-wise and pulse height-wise which is about
-60 mV.

\begin{figure}
\begin{centering}
\includegraphics[width=0.6\textwidth]{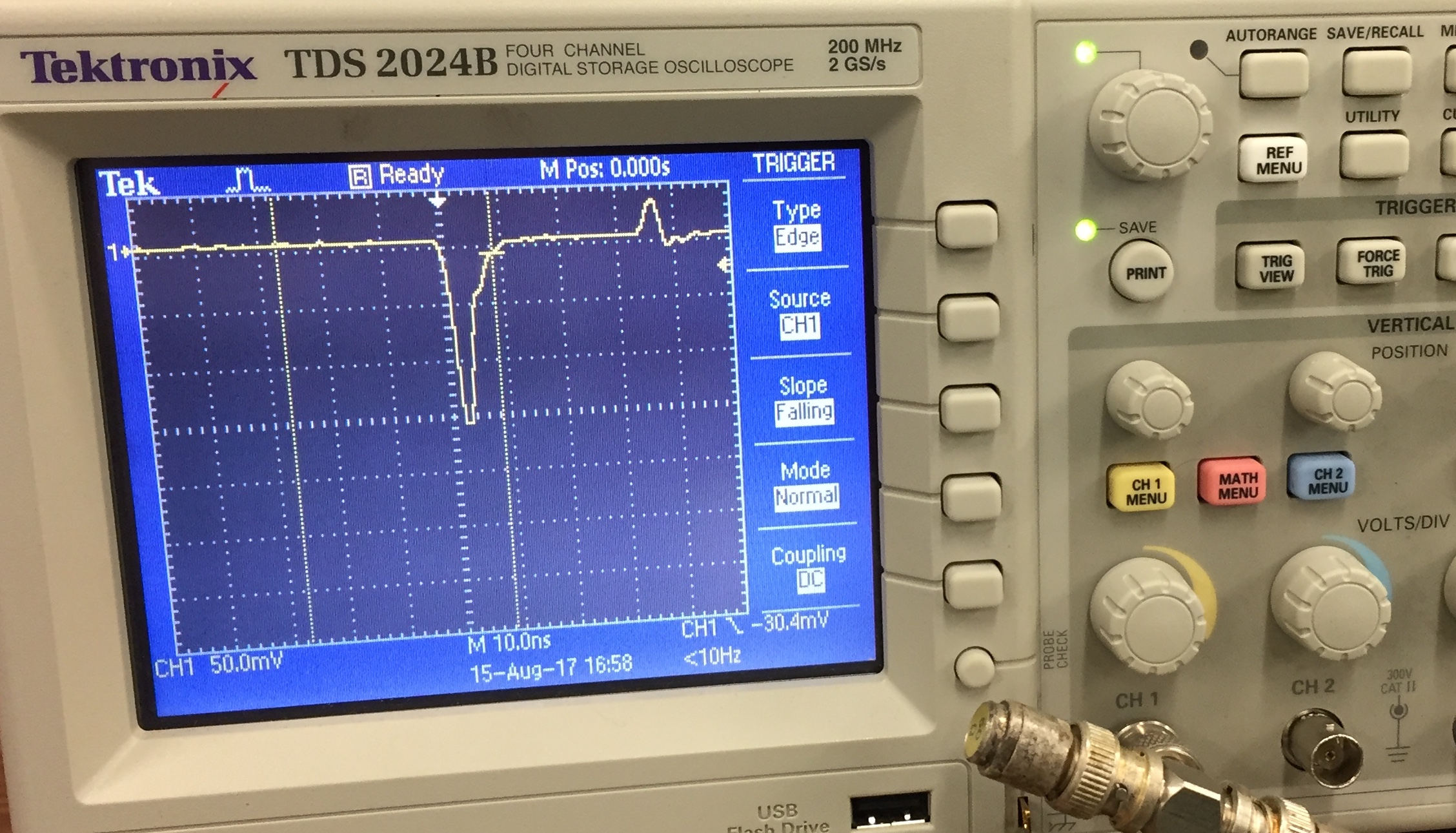}
\par\end{centering}
\caption{Cherenkov light measurement with water in the bin\label{fig:signal-water-1} }
\end{figure}

The expected Cherenkov signal on the other hand has to be rarer and
should have a much higher pulse hight. Such an event was observed
after about 15 mins as shown in Fig.\ref{fig:signal-water-1}. Here
the pulse hight is about -150 mV as expected from Cherenkov light
from a high energy cosmic ray. It is difficult to further elaborate
on the cosmic ray's properties without calibrating the detector in
a test beam. However, it is thought that by improving the quality
of the inner reflecting surfaces, i.e. by collecting more light, the
sensitivity of the setup can be increased. Of course, adding at least
one scintillator to the system and measuring the timing between the
scintillator and Cherenkov would largely improve the setup.

\section{Conclusions and outlook}

Although this particular detector was produced with the minimum of
investment, aiming a proof of the principle, a more developed detector
using not tap water but Gadolinium doped pure water has been submitted
to the Turkish Research and Technical Council, TUBITAK. The aim is
to measure the anti-neutrino flux from the soon-to-be-build nuclear
reactors in Turkey \cite{pub}. This current setup would be extremely
useful for educational and demonstration purposes, if PMT, high voltage
and oscilloscope requirements can be eased. For the first two, one
could consider the utilization of Silicon photo multipliers.
The last one can be matched by a small portable computer running an
oscilloscope and display application. There are such low cost USB
devices \cite{bitscope}and computers capable of running these \cite{raspberry}.
With such a setup it would be possible to display the cosmic rays
and measure their energies in real-time by adapting various materials
with $\epsilon>1$. Furthermore, if batteries can be used as both
low voltage and high voltage power sources, the detector would become
cable-independent, a major step for using it in particle physics education.
\begin{acknowledgments}
The authors would like to thank B.Bilki for useful discussions, helping detector system setup and
a careful reading of the manuscript, and Y.Onel for providing the PMT.
\end{acknowledgments}

\end{document}